\documentclass[twocolumn,showpacs,amsmath,amssymb]{revtex4}

\usepackage{graphicx}
\usepackage{dcolumn}
\usepackage{bm}
\newcommand{\etal}{\emph{et al.}}
\newcommand{\be}{\begin{equation}}
\newcommand{\ee}{\end{equation}}
\newcommand{\bfig}{\begin{figure}}
\newcommand{\efig}{\end{figure}}

\begin{document}
\title{Evidence for massive bulk Dirac Fermions in Pb$_{1-x}$Sn$_x$Se from Nernst and thermopower experiments 
}
\author{Tian Liang$^1$, Quinn Gibson$^2$, Jun Xiong$^1$, Max Hirschberger$^1$, Sunanda P. Koduvayur$^1$, R. J. Cava$^2$ and N. P. Ong$^1$
}
\affiliation{
Department of Physics$^1$ and Department of Chemistry$^2$, Princeton University, Princeton, NJ 08544
}

\date{\today}
\begin{abstract} 
\end{abstract}


\maketitle                   
{Recently, topological surface states (SS) protected by mirror symmetry
were predicted to exist in the rocksalt IV-VI semiconductors. Several groups have 
observed these SS in (Pb,Sn)Te, (Pb,Sn)Se and SnTe using photoemission. 
An underlying assumption in the theory is that
the SS arise from bulk states describable as massive Dirac states,
but this assumption is untested. Here we show that the thermoelectric response of the bulk states 
display features specific to the Dirac spectrum. By relating the carrier density to the peaks in the quantum oscillations,
we show that the $N=0$ Landau Level (LL) is non-degenerate. This finding provides robust evidence that the
bulk states are indeed massive Dirac states. In the
lowest LL, $S_{xx}$ displays a striking linear increase vs. magnetic field characteristic of 
massive Dirac fermions. 
In addition, the Nernst signal displays a sign anomaly in the gap inverted phase at low temperatures.
}

\newpage
The rock-salt IV-VI semiconductors have been identified by Fu and collaborators~\cite{Fu,Hsieh} 
as a novel class of insulators -- the topological crystalline insulators (TCIs) --
which display surface states that are protected by crystalline symmetry. 
The topological surface states in TCIs are to be contrasted with those in the widely investigated Z2 
invariant topological insulators (TIs), which are protected by
time-reversal invariance~\cite{HasanKane,QiZhang}. Angle-resolved photoemission spectroscopy (ARPES) experiments
have obtained evidence for the surface states in Pb$_{1-x}$Sn$_x$Se~\cite{Story}, SnTe~\cite{Takahashi} and Pb$_{1-x}$Sn$_x$Te
~\cite{Hasan}.

\begin{figure*}[t]
\includegraphics[width=15 cm]{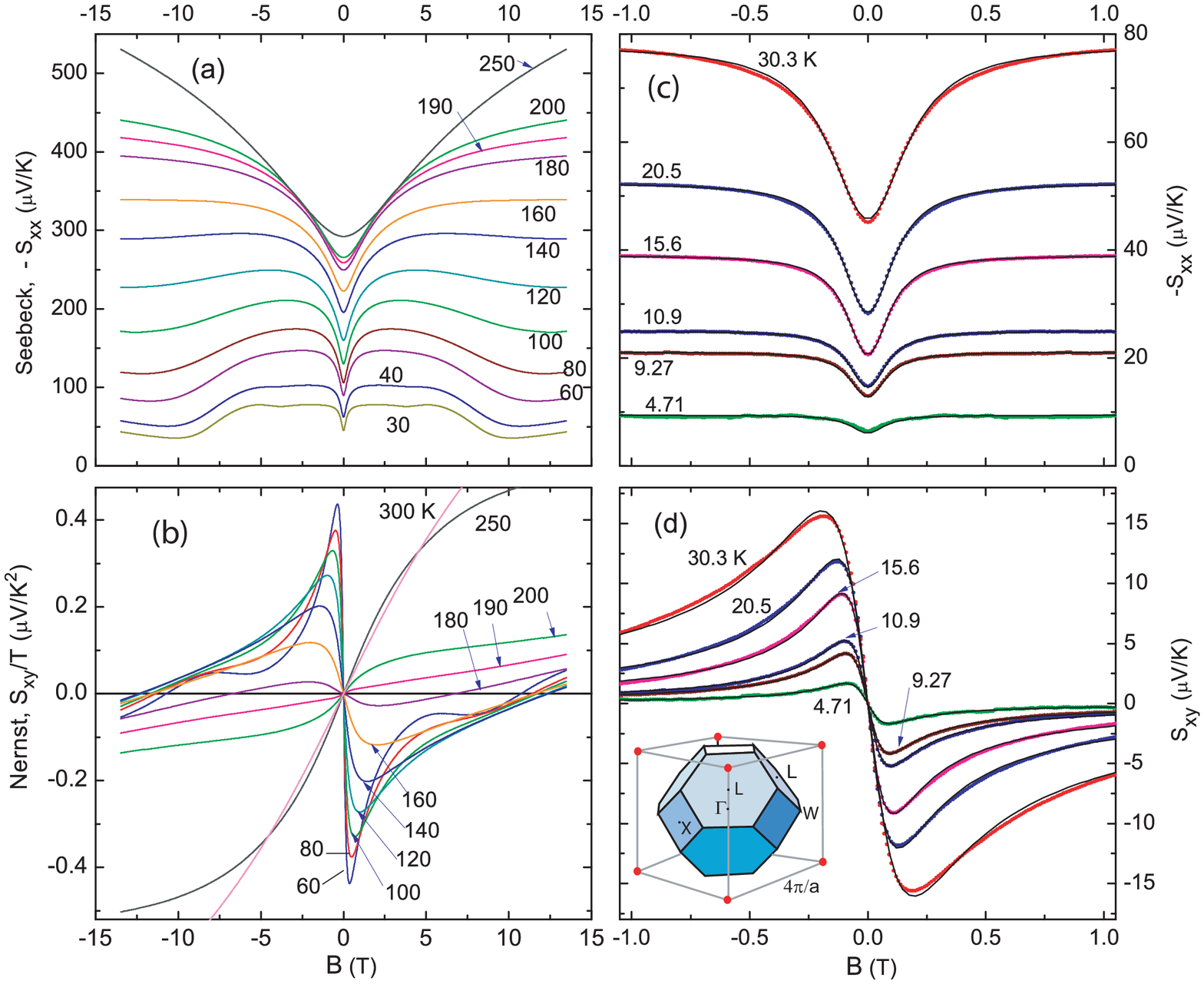}
\caption{\label{figSN} 
The field dependence of the thermopower $S_{xx}$ and Nernst effect $S_{xy}$ in high-mobility Pb$_{1-x}$Sn$_x$Se
($x$ = 0.23). Panel (a) shows curves of $S_{xx}$ vs. $B$ at selected $T$ from 30 to 250 K (sample 1). At each $T$, the $V$-profile 
bracketing $B=0$ reflects the rapid cross-over from small-$\mu B$ to large-$\mu B$ regime.
Panel (b) plots the Nernst signal $S_{xy}/T$ from 60 to 300 K. The sharp peaks reflect the semiclassical response. 
An anomalous sign change occurs at $T_{inv}$ = 180 K. 
Panels (c) and (d) display the fits Eqs. \ref{S} and \ref{N} (thin curves) to $S_{xx}$ and $S_{xy}$ at low $B$.
For $S_{xy}$ (Panel d), we have had to invert the sign.
At 30.3 K, the best-fit values of $\mu$, ${\cal D}$ and ${\cal D}_H$ are 51,404 cm$^2$/Vs,
61.5 eV$^{-1}$ and 104.6 eV$^{-1}$, respectively. At 4.71 K, the corresponding values are 113,250 cm$^2$/Vs,
52.3 eV$^{-1}$ and 81.3 eV$^{-1}$. 
The inset shows the L (111) points on the hexagonal faces of the Brillouin Zone.
}
\end{figure*}

In the alloys Pb$_{1-x}$Sn$_x$Te and Pb$_{1-x}$Sn$_x$Se, the bulk electrons occupy 4 small Fermi Surface (FS) pockets located at
the $L$ points in $\bf k$ space (Inset, Fig. \ref{figSN}). The conduction
band is predominantly derived from the cation Pb $6p$ orbitals whereas the uppermost valence band is predominantly
anion $4p$ (or $5p$) orbitals (ordering similar to the atomic limit)~\cite{Wallis}. 
As the Sn content $x$ increases, the system undergoes gap inversion when $x$ 
exceeds a critical value $x_c$~\cite{Dimmock,Strauss,Dow,Daw}. 
In samples with $x\ge x_c$,
gap inversion occurs when the temperature $T$ is lowered below the 
gap-inversion temperature $T_{inv}$. The ARPES experiments~\cite{Story,Takahashi,Hasan}
confirm that the predicted topological surface states appear in the gap-inverted phase. 

The new topological ideas invite a fresh look at the bulk states of the IV-VI semiconductors.
To date, the gap-inversion appears to have no discernible effect on transport properties (the resistivity, Hall coefficient
and thermopower vary smoothly through $T_{inv}$). This is surprising given that transport probes the
states at the Fermi level. Moreover, a long-standing prediction~\cite{Wallis,Svane} 
is that the bulk electrons occupy states described by the massive Dirac Hamiltonian. 
This assumption underlies the starting Hamiltonian of Hsieh \etal~\cite{Hsieh}. However, no experimental 
test distinguishing the massive Dirac from the Schr\"{o}dinger Hamiltonian has appeared to our knowledge.

We have grown crystals of Pb$_{1-x}$Sn$_x$Se
($x$ = 0.23) in which the $n$-type carriers have high mobilities ($\mu$ = 114,000 cm$^2$/Vs at 4 K). The low
electron density (3.46$\times 10^{17}$ cm$^{-3}$) enables the quantum limit to be reached at
7.7 T (measurements reveal that holes are absent).
In addition to resistivity, we have used both thermopower and the Nernst effect to probe the 
states in fields up to 34 Tesla. Surprisingly, the Nernst signal is observed to change its sign
at $T_{inv}$. To date, this appears to be the only transport or thermodynamic
quantity that is strongly affected by gap inversion.

In a thermal gradient $-\nabla T||\bf\hat{x}$, the diffusion of carriers produces an electric field $\bf E$ which
is expressed as the thermopower signal $S_{xx} = -E_x/|\nabla T|$, and the Nernst signal 
$S_{xy} = E_y/|\nabla T|$.
In the semiclassical regime, the Mott relation~\cite{Ziman} simplifies $S_{xx}$ and $S_{xy}$ to the form (see Methods)
\begin{eqnarray}
S_{xx}(B) &=& {\cal A} \,\left(\frac{\sigma^2}{\sigma^2+\sigma_{xy}^2} {\cal D}+ 
\frac{\sigma_{xy}^2}{\sigma^2+\sigma_{xy}^2} {\cal D}_H \right)			
\label{S} \\
S_{xy}(B) &=& {\cal A} \,\frac{\sigma\sigma_{xy}}{\sigma^2+\sigma_{xy}^2} \left({\cal D}_H - {\cal D} \right),
\label{N}
\end{eqnarray}
where ${\cal A} = \pi^2k_B^2T/3e$. The dependence on $B$ appears only in the conductivity matrix elements $\sigma_{ij}(B)$
(for brevity, we write $\sigma \equiv \sigma_{xx}$).
The parameters ${\cal D} = \partial\ln\sigma/\partial \zeta$ and ${\cal D}_H = \partial\ln\sigma_{xy}/\partial \zeta$
are independent of the mobility $\mu$ ($\zeta$ is the chemical potential). 
Equation \ref{S} describes the cross-over in $S_{xx}$ from ${\cal A}{\cal D}$ 
(at $B=0$) to ${\cal A}{\cal D}_H$ when $\mu B\gg 1$. Correspondingly, $S_{xy}$ increases linearly from 0
to peak at the value $\frac12 {\cal A}({\cal D}_H - {\cal D})$ at $B = 1/\mu$ before falling as $1/B$
when $\mu B\gg 1$. 
For $n$-type carriers, both $\cal A$ and $\sigma_{xy}$ are negative, and $S_{xx} <0$. 
By Eq. \ref{N}, the Nernst signal $S_{xy}$ is \emph{positive} if ${\cal D}_H>{\cal D}$
(we discuss the sign convention in Methods.) In terms of the exponents $\beta$ and $\beta_H$ defined by
$\sigma(E)\sim E^\beta$ and $\sigma_{xy}\sim E^{\beta_H}$, we have ${\cal D} = \beta/E_F$ and ${\cal D}_H = \beta_H/E_F$.

Even for one-band systems, Eqs. \ref{S} and \ref{N} have not received much experimental attention, possibly 
because real materials having only a single band of carriers
with a low density (and high mobility) are rare. The analysis of $S_{ij}$ 
is complicated by the extreme anisotropy of the FS pockets in many semi-metals. 
For recent Nernst measurements on Bi and Bi$_2$Se$_3$, see
Ref. \cite{Behnia1,Behnia2}. For results on $S_{ij}$ in graphene, see Refs. \cite{Kim,Wei,Checkelsky}.
The angular variation of the SdH period in Bi is investigated in Refs. \cite{LuLi,Behnia3}. 

We show that Pb$_{1-x}$Sn$_x$Se crystals that are hole-free satisfy these 3 constraints. 
We find that Eqs. \ref{S} and \ref{N} provide a very good fit to $S_{xx}$ and $S_{xy}$ in the semiclassical regime.
From the fits, we can identify the sign anomaly of the Nernst signal with the gap-inverted phase.

Strong quantum oscillations are observed in both $S_{xx}$ and $S_{xy}$. By analyzing the
LL oscillations, we show that the $N=0$ LL has only one spin degree of freedom
unlike the higher LLs. This provides robust evidence that the carriers are massive 
bulk Dirac fermions, and confirms a premise underlying the starting Hamiltonian in Ref. \cite{Hsieh}. 
In the quantum regime (only $N=0$ LL occupied), we observe an unusual thermoelectric response.
The thermopower increases linearly with $B$ to our maximum $B$ (34 T). We argue that this
profile can be accounted for by the massive Dirac dispersion.

\vspace{5mm}\noindent
{\bf Results}\\\noindent
{\bf Semiclassical regime} Figure \ref{figSN}a plots curves of the thermopower $S_{xx}$ vs. $B$
for selected $T$. From 250 to 160 K, the dominant feature is the 
rapid increase in weak $B$ followed by saturation to
a $B$-independent plateau at large $B$. 
As noted, the Nernst signal (shown as $S_{xy}/T$ in Fig. \ref{figSN}b) changes from positive
to negative as $T$ is decreased below 180 K (identified with $T_{inv}$).

As shown in Fig. \ref{figSN}c, the curves of $S_{xx}$ versus $B$ fit very well to Eq. \ref{S}
in the semiclassical regime ($|B|<$ 1 T). Likewise, below 100 K, the curves of $S_{xy}$ also fit well to Eq. \ref{N}
up to an overall sign (Fig. \ref{figSN}d).  Although the 
fit parameters ($\mu$, ${\cal D}$, ${\cal D}_H$) for $S_{xx}$ are independent of those for $S_{xy}$, 
we find that they agree with each other (at the level of $\pm 2\%$) below 60 K (see Methods). 
At each $T$, the two curves, $S_{xx}(B)$ and $S_{xy}(B)$, are described by just 3 parameters. 
This provides a potent self-consistency check of
Eqs. \ref{S} and \ref{N}. As a further test, we have also fitted the measured conductivity tensor $\sigma_{ij}(B)$ and obtained similar values for $\mu$ 
below 100 K (Methods). By and large, the close fits to both tensors $S_{ij}$ and $\sigma_{ij}$ demonstrate that we have one band of carriers
below 100 K. 

\begin{figure*}[t]
\includegraphics[width=9 cm]{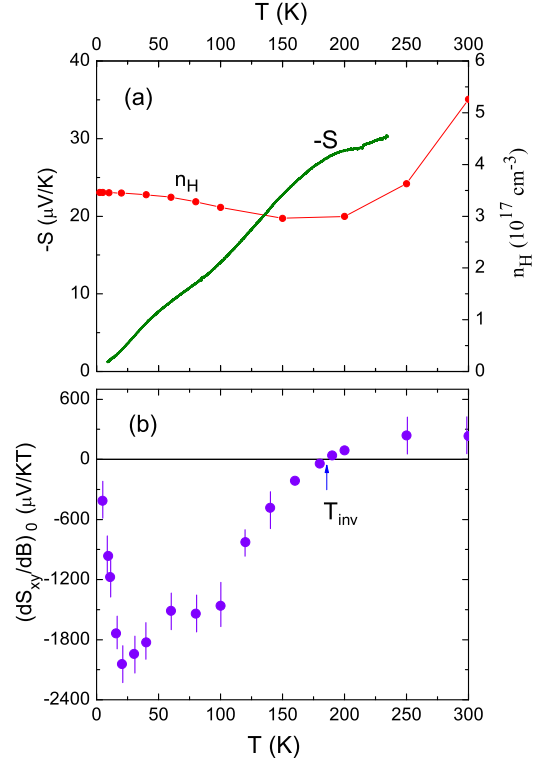}
\caption{\label{figSHallvsT} The temperature dependence of Hall density, 
thermopower and Nernst slope in Pb$_{1-x}$Sn$_x$Se ($x = 0.23$, sample 1).
Panel (a): The $T$ dependence of the Hall density $n_H = B/\rho_{yx}e$ inferred from the Hall resistivity $\rho_{yx}$
and the zero-$B$ thermopower $S(T) \equiv S_{xx}(T,B=0)$ in Pb$_{1-x}$Sn$_x$Se ($x$ = 0.23). The Hall signal is $n$-type at all $T$. 
Below 20 K, $n_H$ equals 3.46$\times 10^{17}$ cm$^{-3}$ (sample 1). $n_H$ increases significantly above 200 K signalling
thermal activation of holes across the band gap.
Panel (b) plots the $T$ dependence of the initial slope of the Nernst signal
$dS_{xy}/dB$ ($B\to 0$) to show the sign change at $T_{inv}$.
}
\end{figure*}

The semiclassical expressions are no longer valid when quantum oscillations appear
at higher $B$. In particular, the giant step at 7.7 Tesla in the curves at 30 and 40 K (Fig. \ref{figSN}a)
is a relic of the quantum regime that remains resolvable up to 100 K. The step plays a key role in the discussion later.
In the opposite extreme above 100 K, the two sets of fit parameters begin to deviate. 
The disagreement is especially acute near 180 K where $S_{xy}$ changes sign. 
We reason that the one-band
assumption breaks down because of strong thermal activation of holes as the gap closes and re-opens across $T_{inv}$.
The evidence comes from the $T$ dependence of the Hall density $n_H = B/e\rho_{yx}$ (solid circles in Fig. \ref{figSHallvsT}a).
Whereas $n_H$ is nearly $T$-independent below 80 K, in agreement with the one-band model, 
it deviates upwards above 180 K. Thermal activation of a large population of holes
leads to partial cancellation of the Hall $E$-field and a reduction in $|\rho_{yx}|$.

In Fig. \ref{figSHallvsT}a, we have also plotted the zero-$H$ thermopower $S\equiv S_{xx}(0)$ 
to bring out its nominally $T$-linear variation below 100 K (bold curve).
The large value of the slope $S(T)/T$ = 1.41 $\mu$V/K$^2$ implies an unusually small $E_F$.

As discussed, the Nernst signal changes sign at $T_{inv}$= 180 K. The $T$ dependence of its initial slope
$dS_{xy}/dB$ ($B\to 0$) is displayed in Fig. \ref{figSHallvsT}b.  
From the fits to Eqs. \ref{S} and \ref{N}, we may address the 
interesting question whether the sign anomaly occurs in the gap-inverted phase ($T<T_{inv}$)
or in the uninverted phase. On both sides of $T_{inv}$, the fits of $S_{xx}$ imply 
${\cal D}_H>{\cal D}$ (i.e. $|S_{xx}|$ always increases as $\mu B$ goes from 0 to values $\gg 1$).
As $S_{xy}\sim ({\cal D}_H -{\cal D})$, we should observe a positive $S_{xy}$. Hence
the sign anomaly occurs in the gap-inverted phase (in Fig. \ref{figSN}d, we multiplied the curves 
by an overall minus sign). The sign of the Nernst signal below $T_{inv}$ disagrees with
that inferred from Eqs. \ref{S} and \ref{N}, despite the close fit.
Further discussion of the sign anomaly is given below (see Discussion). However, we note 
that the sign of $S_{xy}$ is independent of the carrier sign. As seen in Fig. \ref{figSHallvsT}a,
both $S$ and $n_H$ vary smoothly through $T_{inv}$ without a sign change.

\begin{figure*}[t]
\includegraphics[width=9 cm]{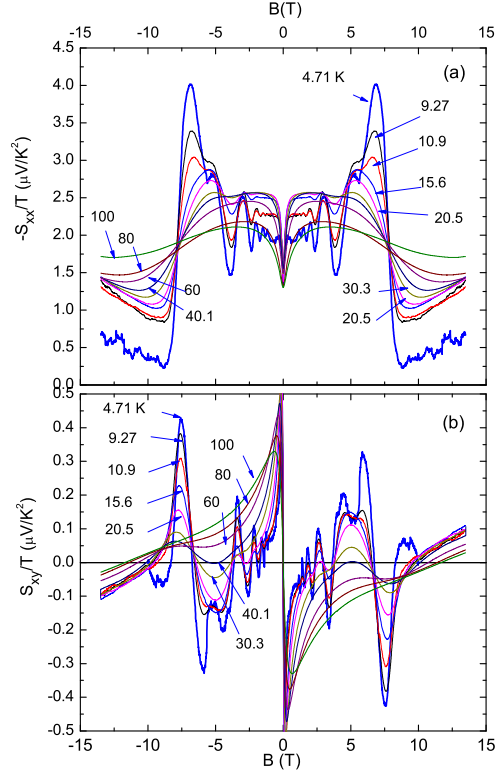}
\caption{\label{figSNlow} 
Quantum oscillations in the thermopower and Nernst signal of Pb$_{1-x}$Sn$_x$Se ($x = 0.23$, sample 1) at temperatures below 100 K.
Panel (a) shows curves of $S_{xx}/T$ vs. $B$ at $T$ = 4.71 to 100 K. $|S_{xx}/T|$ displays a maximum when $\zeta$
is at the DOS maximum in each LL. At 4.71 K, the $N=1$ LL (5-7 T) displays a weak spin-splitting.
The giant step at 7.7 T occurs when $\zeta$ enters the $N=0$ LL.
Panel (b) shows the curves of $S_{xy}/T$ for the same $T$. The sharp resonance-like peaks at low fields 
are the semiclassical response of large-$\mu$ electrons. Below 30 K, they are eclipsed by strong quantum oscillations. 
}
\end{figure*}

\begin{figure*}[t]
\includegraphics[width=7 cm]{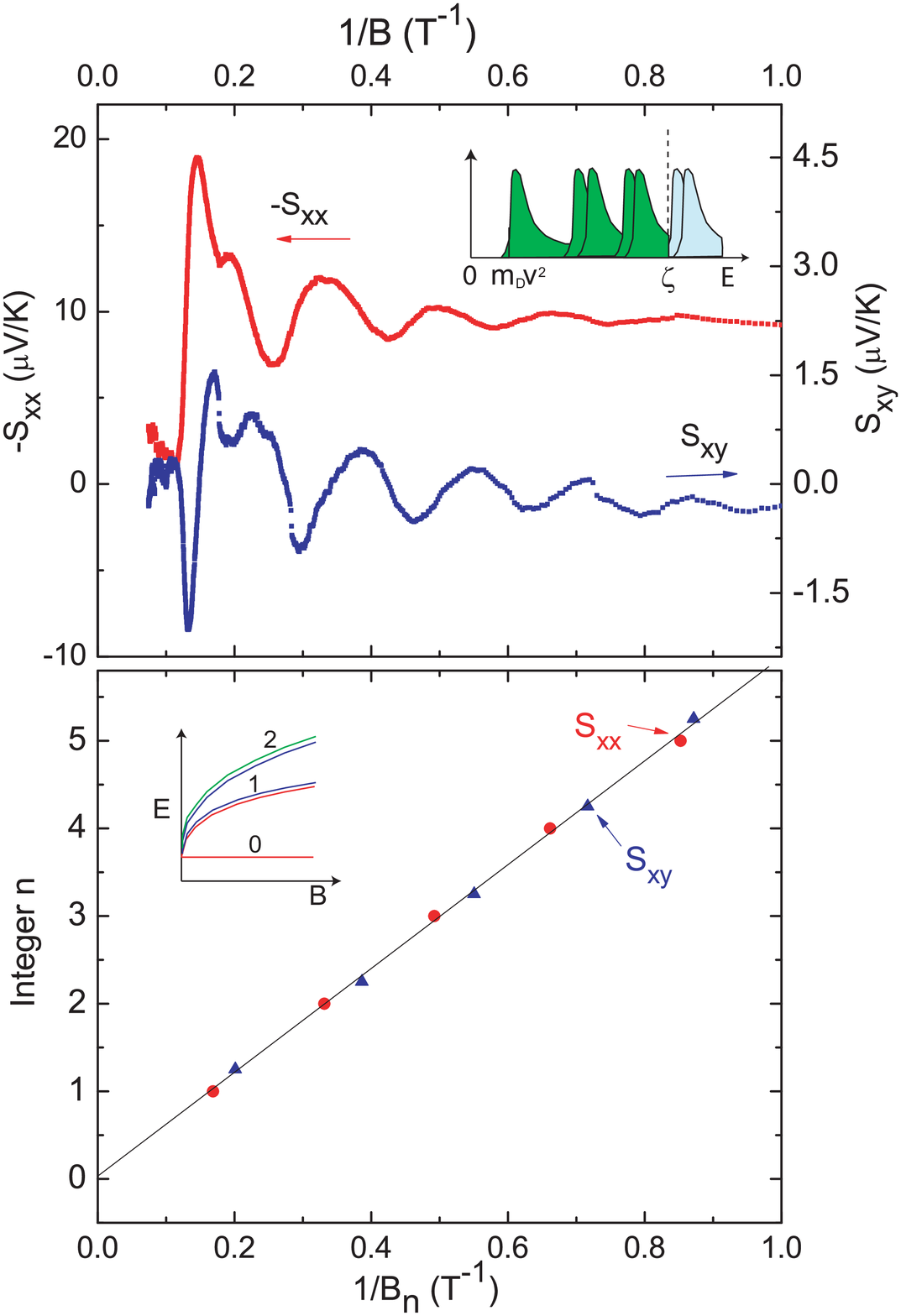}
\caption{\label{figindex} 
Quantum oscillations in Pb$_{1-x}$Sn$_x$Se ($x = 0.23$). Panel (a) compares curves of $S_{xx}$ and $S_{xy}$ vs. $1/B$ 
at 4.71 K. The maxima in $S_{xy}$ are shifted by a $\frac14$ period relative to the maxima in $S_{xx}$. 
The $N=1$ LL shows a weak spin-splitting. The sketch (inset) shows the peaks in the DOS of each LL for 3D massive Dirac fermions.
Panel (b) displays the index plot of $B_n$ corresponding to the maxima in $|S_{xx}|$ (solid circles)
and $S_{xy}$ (triangles) versus the integers $n$. The straight line is the relation
${\cal S}_F = 2\pi(n+\gamma)/\ell_B^2$ where ${\cal S}_F$ is the FS section and $\gamma$ the Onsager phase.
The maxima in $S_{xy}$ are shifted by $\frac14$ in $n$. 
From the slope we infer the Fermi wavevector $k_F$ = 0.0134 \AA$^{-1}$ and $n_e$ = 8.20$\times 10^{16}$ cm$^{-3}$ (per valley). 
The inset shows the LL energy $E$ vs. $B$ in the massive Dirac spectrum for $k_z=0$ (Ref. \cite{Wolff}). 
}
\end{figure*}

\vspace{3mm}\noindent
{\bf Quantum oscillations}
As shown in Fig. \ref{figSNlow}, oscillations in $S_{xx}$ and $S_{xy}$
grow rapidly below 60 K to dominate the weak-$B$ semiclassical profile. 
The most prominent feature in $S_{xx}$ is the large step-decrease
at the field $B_1$ = 7.7 T (at which the chemical potential $\zeta$ jumps from the $N=1$ LL to the $N=0$ LL).
In the Nernst curves, plotted as $S_{xy}/T$ in Panel (b), the quantum oscillations are more
sharply resolved. Because $S_{xy}$ is the off-diagonal term of the tensor $S_{ij}$, its
maxima (or minima) are shifted by $\frac14$ period 
relative to the extrema of the diagonal $S_{xx}$ (analogous to the shift of $\sigma_{xy}$ relative to $\sigma$). 
This shift is confirmed in Fig. \ref{figindex}a which plots the traces of $S_{xx}$ and $S_{xy}$ vs.
$1/B$. For the analysis below, we ignore the weak spin-splitting 
which is resolved in the $N$=1 LL (and barely in $N=2$).

Figure \ref{figindex}b shows the index plot of $1/B_n$ (inferred from the
maxima in $|S_{xx}|$ and $S_{xy}$) plotted versus the integers $n$. 
From the slope of the line, we derive the Fermi Surface section 
$S_F$ = 5.95 T = 5.67$\times 10^{16}$ m$^{-2}$. Assuming a circular cross-section, we have $k_F$ = 0.0134 \AA$^{-1}$.
The electron concentration per FS pocket is then $n_e = k_F^3/3\pi^2$ = 8.2$\times 10^{16}$ cm$^{-3}$.
As there are 4 pockets, the total carrier density is $4n_e = 3.28\times 10^{17}$ cm$^{-3}$, in good
agreement with the Hall density $n_H$ at 4 K (3.46$\times 10^{17}$ cm$^{-3}$).

\begin{figure*}[t]
\includegraphics[width=9 cm]{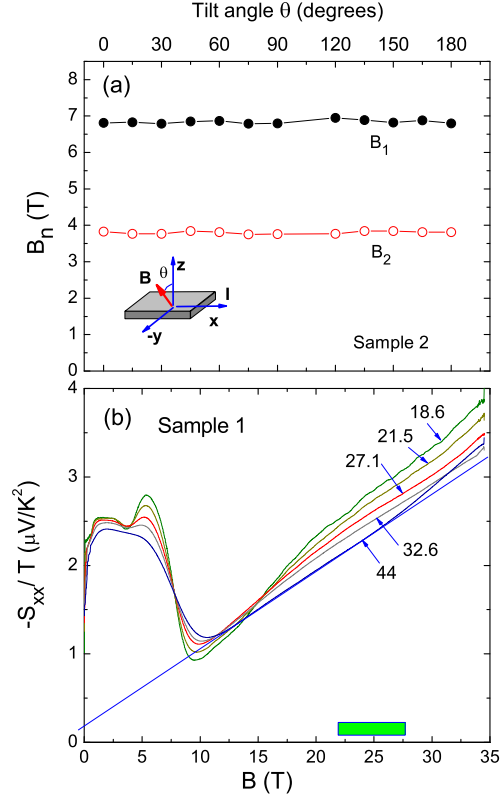}
\caption{\label{fig35T} 
Dependence of SdH period on field-tilt angle and high-field thermopower.
Panel (a): The dependence of the transition fields $B_1$ and $B_2$ versus tilt angle $\theta$ of $\bf B$ in sample 2 ($x=0.23$)
inferred from magnetoresistance ($B_1$ and $B_2$ are the fields at which
$\zeta$ jumps from LL with $N = 1\to 0$ and $N=2\to 1$, respectively).
Within our resolution, no angular dependence of $B_1$ and $B_2$ is observed. $\bf B$ is rotated in
the $y$-$z$ plane (sketch in inset).
Panel (b): High-field measurements of $S_{xx}/T$ to 34 T at several $T$ in sample 1. The $B$-linear dependence smoothly extends through
the region 22-28 T (bar) where the transition $(0,-)\to (0,+)$ should have appeared.
}
\end{figure*}

Using sample 2, we have tracked the variation of the SdH period versus the tilt angle $\theta$ of $\bf B$.
Figure \ref{fig35T}a plots the fields $B_1$ and $B_2$ 
versus $\theta$ ($\bf B$ is rotated in the $y$-$z$ plane). Here, $B_1$ and $B_2$ are the
fields at which $\zeta$ jumps from $N=1\to 0$ and from $N=2\to 1$, respectively.
To our resolution, the SdH period is nearly isotropic. 
The fields $B_1$ and $B_2$ are also independent of tilt angle when $\bf B$ is rotated in the $x$-$y$ plane.
This justifies treating the FS pockets as nominally spherical.

\vspace{3mm}\noindent
{\bf The N=0 Landau Level}
We next address the question whether the bulk states in the inverted phase are Dirac fermions or Schr\"{o}dinger electrons.
The two cases differ by a distinctive feature in their LL spectrum that is robust against
small perturbations. In the quantum limit, the massive Dirac Hamiltonian exhibits
an interesting two-fold difference in degeneracy between the $N=0$ and $N=1$ levels.
Wolff~\cite{Wolff} considered a 3D massive Dirac Hamiltonian with spin-orbit interaction but no Zeeman energy term.
More recently, Serajedh, Wu and Phillips~\cite{Phillips} included the Zeeman energy term as well as a Rashba term
in the massive 2D Dirac Hamiltonian. Other 3D massive Dirac cases are discussed by Bernevig~\cite{Bernevig}. All these
authors find that the $N=0$ LL is non-degenerate with respect to spin degrees, whereas the LLs 
with $N\ne 0$ are doubly spin-degenerate.
(We discuss in Methods a pedagogical example which shows that this anomaly is related to the conservation of states.)
By contrast, for the Schr\"{o}dinger case, all LLs are doubly degenerate.

In Pb$_{1-x}$Sn$_x$Se, the ability to measure accurately both $n_e$ and the ``jump'' field $B_1$ 
provides a crisp confirmation of this prediction.

The energy of the $N^{th}$ LL is $E(N,k_z) = \sqrt{(m_Dv^2)^2 + (\sqrt{2N}\hbar v/\ell_B)^2 + (\hbar vk_z)^2}$,
where $m_D$ is the Dirac mass and $\ell_B = \sqrt{\hbar/eB}$ the magnetic length~\cite{Bernevig}.
At $B_1$, $E_F$ lies just below the bottom of the $N=1$ LL so that all the electrons are accomodated 
in the $N=0$ LL. Integrating the density of states (DOS) for one spin polarization in the 
$N=0$ LL from $m_Dv^2$ to $E_F$, we find (see Methods)
\be
n_{e\uparrow} = \sqrt{2}/(2\pi^2\ell_B^3).
\label{ne1}
\ee
Ignoring the small spin splitting, we equate $B_1$ with 7.7 T.
Equation \ref{ne1} then gives $n_{e\uparrow} = \,9.0\times 10^{16}$ cm$^{-3}$, which agrees within
$10\%$ with the measured $n_e$ (the agreement is improved if we correct for spin splitting). 
All the electrons are accomodated by an $N=0$ LL that is non-degenerate, in agreement with the prediction
for massive Dirac states~\cite{Wolff,Phillips,Bernevig}, but disagreeing with the Schr\"{o}dinger case by a factor of 2.
Since the singular spin degeneracy of the $N=0$ LL cannot be converted to a double degeneracy,
the experiment uncovers a topological feature of the bulk states that is robust.  
As predicted in Refs. \cite{Wolff,Phillips,Bernevig}, the $N=0$ LL has only one spin state
(0,+); the spin-down partner (0,-) is absent.

To check this further, we extended measurements of $S_{xx}$ to 34 T to search for the transition 
from the sublevel (0,-) to (0,+) (which should occur if the $N=0$ LL were doubly degenerate). From extrapolation of the spin-split $N=1$ and $N=2$ LLs, 
we estimate that the transition (0,-)$\to$(0,+) should appear in the interval 22-28 T. As shown 
in Fig. \ref{fig35T}b, the measured curves show no evidence for this transition to fields up to 34 T. 

Finally, we note an interesting thermopower feature in the quantum limit.
At fields above $B_1$, 
$S_{xx}$ displays a $B$-linear profile that extends to 34 T (Fig. \ref{fig35T}). 
The $B$-linear behavior is most evident in the curve at 44 K. As $T$ is decreased to 18.6 K,
we resolve a slight downwards deviation from the linear profile in the field interval 10-20 T.
The $B$-linear profile appears to be a characteristic property of massive Dirac fermions in the 
quantum limit. 
We discuss below a heuristic, semiclassical approach that reproduces the observed profile.

\vspace{5mm}\noindent
{\bf Discussion}\\
\noindent
We summarize the electronic parameters inferred from our experiment and relate
them to ARPES measurements.

As noted, the FS section derived from the index plot (Fig. \ref{figindex}b) corresponds
to a total electron density $4n_e = 3.28\times 10^{17}$ cm$^{-3}$, in good
agreement with the Hall density $n_H$ at 4 K (3.46$\times 10^{17}$ cm$^{-3}$).

We may estimate $E_F$ from the slope of the thermopower
$S(T)/T$ = 1.41 $\mu$V/K$^2$.
Using the Mott expression $S(T) = (\pi^2/3)(k_B/e)(k_BT/E_F)\beta$, we find for the Fermi energy 
$E_F = 17.0\;\beta$ meV. For the massive Dirac dispersion, we have $n_e\sim k_F^3$, which implies
that $\beta$ has the minimum value 3 (if the mobility increases with $E$, $\beta$ is larger).
Using the lower bound, $\beta = 3$, $S/T$ gives $E_F $ = 51 meV. 

These numbers may be compared with ARPES. We estimate the Fermi velocity from the expression 
$v \simeq E_F/\hbar k_F$ (valid when $E_F\gg m_Dv^2$ with $m_D$ the Dirac mass). Using our values of $E_F$ and $k_F$,
we find $v$ = 5.74$\times 10^5$ m/s as the lower bound. 
Although ARPES cannot resolve $v$ in the conduction band, the best
ARPES estimate~\cite{Hasan2} for the hole band velocity is 5.6$\times 10^5$ m/s, in good agreement
with our lower bound. It is likely that the conduction band has a higher velocity (which would then require
$\beta >$3).

\vspace{3mm}\noindent
One of our findings is that gap inversion changes the sign of the Nernst signal.
Because the energies of states involved in gap inversion are very small, the resulting dispersion
can be hard to resolve by ARPES~\cite{Hasan2}. Transport quantities would appear
to be more sensitive to these changes. As noted, however, most transport quantities are either unaffected or only mildly perturbed.
The Hall effect and thermopower are unchanged in sign across $T_{inv}$ (Fig. \ref{figSHallvsT}a). While $n_H$ shows a gradual increase,
this is largely attributed to thermal activation of holes across a reduced gap for $T>T_{inv}$.
Hence, the dramatic sign-change observed in $S_{xy}$ stands out prominently; its qualitative nature may provide
a vital clue. 

It has long been known~\cite{Strauss} that, in the lead rock-salt IV-VI semiconductors, the energy gap $E_g$ 
undergoes inversion as the Sn content $x$ increases from 0. Moreover, within a narrow range of $x$, 
gap inversion is also driven by cooling a sample (the critical temperature is $x$ dependent within this interval).
Strauss~\cite{Strauss} performed early optical transmission measurements of $E_g$ 
in a series of single-crystal films of Pb$_{1-x}$Sn$_x$Se with $x$ ranging from 0 to 0.35. 
For $x=0.25$, he reported that $E_g$ closes at 195 K. A slight interpolation of his data shows that, at our doping $x=0.23$, $E_g$ 
should vanish at 179 K, remarkably close to our $T_{inv}$ = 180 K. The recent ARPES measurements of Dziawa \etal~\cite{Story}
is consistent with $E_g$ closing between 100 and 200 K. ARPES measurements by Hasan's group~\cite{Hasan2}
on a crystal from the same batch as our samples shows that $E_g$ crosses zero between 80 and 150 K.
Given the ARPES resolution, these results are all consistent with our inference that 
our $T_{inv}$ corresponds to the gap inversion temperature.
Hence we reason that the Nernst signal changes sign either at, or very close to, the gap inversion temperature.
The inverted sign of $S_{xy}$ below 180 K in Figs. \ref{figSN}c,d occurs in the gap inverted phase.
(We refrain from making the larger claim that this is also the topological transition
because we are unable to resolve the surface states in our experiments.)

The fits of $S_{xy}$ to Eq. \ref{N} (Fig. \ref{figSN}d) shows that the curves below 100 K
are well-described by the Boltzmann-Mott expression assuming a single band of carriers, but there is an
overall sign disagreement. Despite the sign problem, the analysis singles out the physical factors 
that fix the sign, and delineates the scope of the problem. 
For example, reversing the sign of both $\beta$ and $\beta_H$ inverts the sign of
$S_{xy}$, but also that of $S_{xx}$. Alternately, one might try 
reversing the signs of $\beta$, $\beta_H$ and $e$ simultaneously. This will invert the sign of $S_{xy}$ 
but leave $S_{xx}$ unchanged. However, $\rho_{yx}$ is forced to change sign.

The analysis assumes that, in the gap-inverted phase, the FS is simply connected. 
This may not be valid. Gap inverion may lead to the existence of a small pocket 
surrounded by a larger FS sheet (topologically similar to the FS of the
``giant Rashba'' material BiTeI~\cite{BTE}). As the small pocket dominates the thermoelectric response,
the Nernst effect may be detecting this novel situation.
These issues will be left for future experiments.

\vspace{3mm}\noindent
We may attempt to understand the striking $B$-linear profile of $S_{xx}/T$ in Fig. \ref{fig35T} using
a semiclassical approach. In the $N=0$ LL, the long-lived quasiparticles complete a large number of cyclotron orbits
between scattering events (e.g. from $\mu B\sim$ 220, we estimate this number is $\sim$35 at 20 T).
The scattering results in the drift of the orbit centers $\bf X$ in a direction transverse to the applied $-\nabla T$.
Ignoring the fast cyclotron motion, we may apply the Boltzmann equation to $\bf X$. The thermopower
is then given by the high-$B$ limit of Eq. \ref{S}, $S_{xx}(T,H) \to {\cal A}\beta_H'/E_F$,
where $E_F$ is now measured from the bottom of the $N=0$ LL, and
$\beta_H'$ differs from the weak-field $\beta_H$. In this picture, the $B$ dependence of $S_{xx}$ arises solely from how
$E_F$ changes with $B$.

For $B>B_1$, only the $N=0$ LL is occupied. From Eq. \ref{ne0} (Methods), we have the relation
between $E_F, \, n_e$ and $B$, viz.
\be
E_F^2 = (m_Dv^2)^2 + \frac{{\cal P}^2}{B^2}, \quad \left({\cal P} = \frac{2\pi^2\hbar^2 v n_e}{g_se} \right).
\label{DiracE}
\ee

In the limit $E_F\gg m_Dv^2$, we obtain the relation $E_F \sim 1/B$. This immediately implies
that $S_{xx}/T$ increases linearly with $B$ as observed. Setting $g_s=1$, we derive from Eq. \ref{DiracE} 
the rate of increase
\be
\frac{\partial S_{xx}/T}{\partial(B)} = \frac{k_B^2}{6\hbar^2} \frac{\beta_H'}{vn_e}.
\label{dSdB}
\ee
Repeating this calculation for the Schr\"{o}dinger case, we get instead $S_{xx}/T\sim B^2$.

From Fig. \ref{fig35T}, the thermopower slope 
$\partial (S_{xx}/T)/\partial B$ = 8.71$\times 10^{-8}$ V/K$^2$T. 
Using the above values of $v$ and $n_e$ in Eq. \ref{dSdB}, 
we find $\partial (S_{xx}/T)/\partial B = 6.1 \beta_H'\times 10^{-8}$ V/K$^2$T. 
The value of $\beta_H'$ is not known. Comparison of the calculated slope with experiment suggests
$\beta_H'\sim$ 1.5. Hence this back-of-the-envelop estimate can account for the
rate at which $S_{xx}/T$ increases with $B$.


\vspace{8mm}\noindent
{\bf Methods}\\
{\bf Semiclassical fits to $\mathbf{S_{xx}}$ and $\mathbf{S_{xy}}$}
In the presence of a magnetic field $\bf B$, an electric field $\bf E$ and a temperature gradient $-\nabla T$ (in an infinite medium), 
the total current density is given by~\cite{Ziman}
${\bf J} = \boldsymbol{\sigma}\cdot {\bf E} +  \boldsymbol{\alpha}\cdot(-\nabla T)$.
Here $\sigma_{ij}$ is the conductivity tensor and $\alpha_{ij}$ is the thermoelectric tensor.
Setting $\bf J$ = 0 (for a finite sample), and solving for $\bf E$, we have
$\bf E = -\boldsymbol{\rho}\cdot\boldsymbol{\alpha}\cdot(-\nabla T)$, with $\boldsymbol{\rho} = \boldsymbol{\sigma}^{-1}$
the resistivity tensor.

In the geometry with $\bf B||\hat{z}$ and $-\nabla T||\bf \hat{x}$, the components of the $E$-field (for an isotropic system) are
\begin{eqnarray}
E_x/|\nabla T| &=& -(\rho_{xx}\alpha_{xx} + \rho_{yx}\alpha_{xy}) \label{ex} \\
E_y/|\nabla T| &=& \rho_{xx}\alpha_{xy} - \rho_{yx}\alpha_{xx}.\label{ey}
\end{eqnarray}
The thermoelectric tensor $S_{ij}$ is given by $E_i = S_{ij}\partial_j T$
($S_{xx}>0$ for hole carriers and $S_{xy}>0$ if $E_y>0$ when $H_z>0$).

The Mott relation~\cite{Ziman}, 
\be
\alpha_{ij} = {\cal A} \left[\frac{\partial\sigma_{ij}}{\partial \varepsilon}\right]_{\zeta}, \quad 
\left({\cal A} = \frac{\pi^2}{3}\frac{k_B^2T}{e}\right),
\label{Mott}
\ee
($k_B$ is Boltzmann's constant, $e$ is the elemental charge and $\zeta$ the chemical potential) 
has been shown to hold under general conditions, e.g. in the Quantum Hall Effect (QHE)
~\cite{Girvin,Jonson}. 
Using Eq. \ref{Mott}, Eqs. \ref{ex} and \ref{ey} reduce to Eqs. \ref{S} and \ref{N}, respectively.

The fits of $S_{ij}$ to these equations displayed in Fig. \ref{figSN}d
were carried out using the one-band, Boltzmann-Drude expressions for the conductivity tensor, viz.
\begin{eqnarray}
\sigma_{xx}(B) &=& N_ee\mu/(1+\mu^2B^2), \label{sxx}\\ 
\sigma_{xy}(B) &=& N_ee\mu^2B/(1+\mu^2B^2),
\label{sxy}
\end{eqnarray}
where the total carrier density $N_e$ is 4$n_e$ ($n_e$ is the density in each of the Fermi Surface 
pocket at the L points).

In the geometry with $\bf B||\hat{z}$ and $-\nabla T||{\bf\hat{x}}$,
we define the sign of the Nernst signal to be that of the $y$-component of the E-field $E_y$. More generally,
if ${\bf E}_N$ is the E-field produced by the Nernst effect,
the sign of the Nersnt signal is that of the triple product ${\bf E}_N\cdot{\bf B\times}(-\nabla T)$.
This agrees with the old convention based on ``Amperean current''~\cite{Bridgman} and 
with the one adopted for vortex flow in superconductors~\cite{Yayu06}.

\begin{figure*}[t]
\includegraphics[width=11 cm]{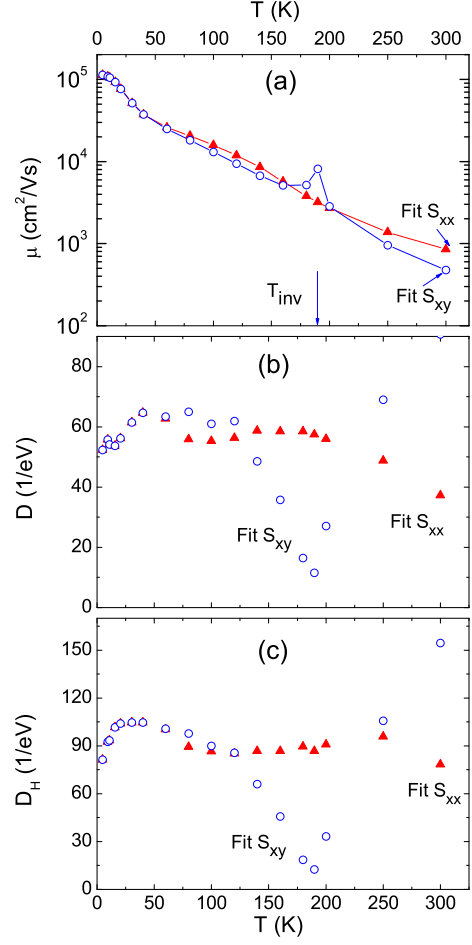}
\caption{\label{figFitParam} 
Variation of fit parameters versus temperature.
The $T$ dependence of the parameters $\mu$ (Panel a), ${\cal D}$ (b) and ${\cal D}_H$ (c) obtained from best fits to the
curves of $S_{xx}$ (solid triangles) and $S_{xy}$ (open circles). The fits are most reliable below 100 K where the values 
obtained from fitting $S_{xx}$ and $S_{xy}$ are in agreement. Above 100 K, disagreement 
between the two sets is significant, especially close to $T_{inv}$ = 180 K. Above 200 K, the one-band model is 
no longer valid.
}
\end{figure*}

At each $T$, we have fitted the measured curves of $S_{xx}$ and $S_{xy}$ vs. $B$ to Eqs. \ref{S} and \ref{N}
using Eqs. \ref{sxx} and \ref{sxy} for the conductivity tensor. The separate fits of $S_{xx}$ and $S_{xy}$
yield two sets of the parameters $\mu$, ${\cal D}$ and ${\cal D}_H$ which are displayed in Fig. \ref{figFitParam}
(solid triangles and open circles, respectively). The 3-parameter fit places strong constraints on the 
curves of $S_{xx}$ and $S_{xy}$. Disagreement between the 2 sets signals that the one-band model is inadequate. 

Below 100 K, the two sets agree well, whereas closer to $T_{inv}$
they begin to deviate. The reason is that Eq. \ref{N} cannot account for the change of sign in the Nernst signal
given the relative magnitudes of ${\cal D}$ and ${\cal D}_H$ fixed by the curves of $S_{xx}$.
Above 200 K, the 2 sets are inconsistent because thermal excitations of holes across the small band-gap is
important at elevated $T$, and the one-band assumption becomes inadequate. 
This is evident in the onset above 200 K of significant $T$ dependence in the Hall density $n_H$ (see Fig. \ref{figSHallvsT}a).

[We remark that $S_{xx} = V_x/\delta T$ is directly obtained from the
observed voltage difference $V_x$ and the temperature difference $\delta T$
between longitudinal electrical contacts (their spatial separation $L_x$ is immaterial). However, for the
Nernst signal, we have $S_{xy} = (V_y/\delta T)(L_x/L_y)$, where $L_y$ is the spatial separation
between the transverse contacts. Hence the aspect ratio $L_y/L_x$ is needed to convert
the observed Nernst voltage $V_y$ to $S_{xy}$. The ratio $L_y/L_x$ is measured to be 4$\pm$0.4. The fits are
improved significantly if this value is refined to 4.20, which we adopt for the curves at all $T$.]

\begin{figure*}[t]
\includegraphics[width=9 cm]{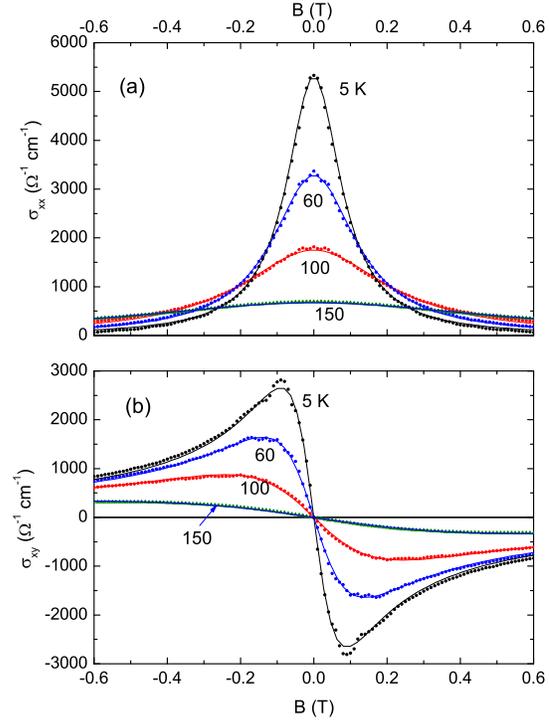}
\caption{\label{figCond} 
Fits of the conductivity tensor of Pb$_{1-x}$Sn$_x$Se ($x=0.23$).
The measured weak-$B$ conductivity $\sigma_{xx}$ (Panel (a)) and Hall conductivity $\sigma_{xy}$ (Panel (b)) of Pb$_{1-x}$Sn$_x$Se ($x=0.23$) vs. $B$ 
at selected $T$ from 5 to 150 K, together with the fits to Eqs. \ref {sxx} and \ref{sxy} (thin solid curves).
}
\end{figure*}

Fits to Eqs. \ref{sxx} and \ref{sxy} of the conductivity tensor measured in the same sample are shown in Fig. \ref{figCond} for weak $B$
at selected $T$ from 5 to 150 K. 
The fits yield values of the mobility $\mu$ similar to those shown in Fig. \ref{figFitParam}a. The 
inferred carrier density $N_e$ is also similar to the measured Hall density $n_H$.

\vspace{3mm}\noindent
{\bf  Indexing the Quantum oscillations}
For 3D systems, one identifies the index field $B_n$ as the field at which the DOS displays a sharp maximum
(diverging as $[E-(n+\frac12)\hbar\omega_c]^{-\frac12}$ in the absence of disorder). From the quantization rule
for areas in $k$-space, $B_n$ is related to the FS cross-section $S_F$ as 
\be
{\cal S}_F = \frac{2\pi}{\ell_B^2}(n+\gamma),
\label{SF}
\ee
where $\ell_B = \sqrt{\hbar/eB}$ and $\gamma$ (the Onsager phase) is $\frac12$ for Sch\"{o}dinger electrons.
The plot in Fig. 3b follows Eq. \ref{SF}. From its slope, we obtain ${\cal S}_F$. The intercept $\gamma$ is 
close to zero in Fig. 3b. We will discuss $\gamma$ elsewhere.

(We note that, in 2D systems in the QHE regime, the index field is the field at which the chemical potential $\zeta$ falls between adjacent LLs,
where the DOS \emph{vanishes}, and the Hall conductance displays a plateau. The difference between 2D and 3D
systems arises because the integer $n$ counts the number of edge states in the QHE case, whereas 
$n$ indexes the DOS peaks in the 3D case. One needs to keep this in mind in interpreting $\gamma$.)

We have verified that the slope in Fig. \ref{figindex}b is insensitive to the tilt angle $\theta$ of $\bf B$ relative to the crystalline axes.
As shown in Fig. \ref{fig35T}a, the SdH period is virtually independent of $\theta$ within the experimental uncertainties,
consistent with negligible anisotropy in the small FS pockets. The good agreement between ${\cal S}_F$
and $n_H$ (Hall density) at 5 K is also evidence for a negligible anisotropy.

\vspace{3mm}\noindent
{\bf  Spin degeneracy in $N$=0 LL}
Knowledge of the field $B_1$ (the transition from $N=1$ to the $N=0$ LL)
and the electron density per valley $n_e$ suffices to determine the spin degeneracy of the $N=0$ LL.

For the 3D Dirac case~\cite{Bernevig}, the energy in the $N^{th}$ LL is 
\be
E_{N,kz} = \sqrt{(m_Dv^2)^2 + (\frac{\sqrt{2N}\hbar v}{\ell_B})^2 + (\hbar v k_z)^2},
\label{Dirac}
\ee
with $m_D$ the Dirac mass, $k_z$ the component of $\bf k$ along $\bf B$, 
and $\ell_B = \sqrt{\hbar/eB}$ the magnetic length.

For the $N=0$ LL, we solve for $k_z(E)$
\be
k_z(E) = \sqrt{E^2-E_{00}^2}/\hbar v,
\label{kz}
\ee
where $E_{00} = m_Dv^2$.

Let us assume that only the $N=0$ LL is occupied. To obtain the relation linking $E_F,\, B$ and $n_e$, we integrate 
the 3D density of states ${\cal D}(E)dE = (g_Lg_s/\pi) dk_z$, with
$g_s$ the spin degeneracy and $g_L = 1/2\pi\ell_B^2$ the 2D LL degeneracy per spin. 
Using Eq. \ref{kz}, we have
\be 
n_e = \int^{E_F}_{E_{00}}{\cal D}(E) dE = \frac{g_Lg_s}{\pi\hbar v} \sqrt{E_F^2-E_{00}^2}.
\label{ne0}
\ee

This equation is valid until $B$ is reduced to the jump field $B_1$, whereafter electrons enter the $N=1$ LL.
At the jump field, $E_F$ lies just below the bottom of the $N=1$ LL, i.e.
$E_F^2 = E_{10}^2 = (m_Dv^2)^2 + (\sqrt{2}\hbar v/\ell_B)^2$.
Using this in Eq. \ref{ne0}, we have 
\be
n_e = \frac{\sqrt{2}g_s}{2\pi^2\ell_B^3} \quad (B=B_1).
\label{ne}
\ee

In relation to Eq. \ref{ne1}, we showed that Eq. \ref{ne} gives a value equal (within 10$\%$) to the total
electron density per valley if $g_s=1$, i.e. when $B>B_1$, all the electrons can be accomodated by the $N=0$ LL
with only one spin polarization. This is direct evidence for the non-degeneracy of
the $N=0$ LL.

Interestingly, Eq. \ref{ne} is identical for the isotropic Schr\"{o}dinger case, for which
\be
E_{N,kz}  = (N+\frac12)\hbar\omega_c + \frac{\hbar^2 k_z^2}{2m},
\label{ES}
\ee
where $\omega_c = eB/m$ and $m$ is the mass.
However, for the $N=0$ LL of the Schr\"{o}dinger spectrum, we must have $g_s = 2$, so it can be excluded.

\begin{figure*}[t]
\includegraphics[width=7 cm]{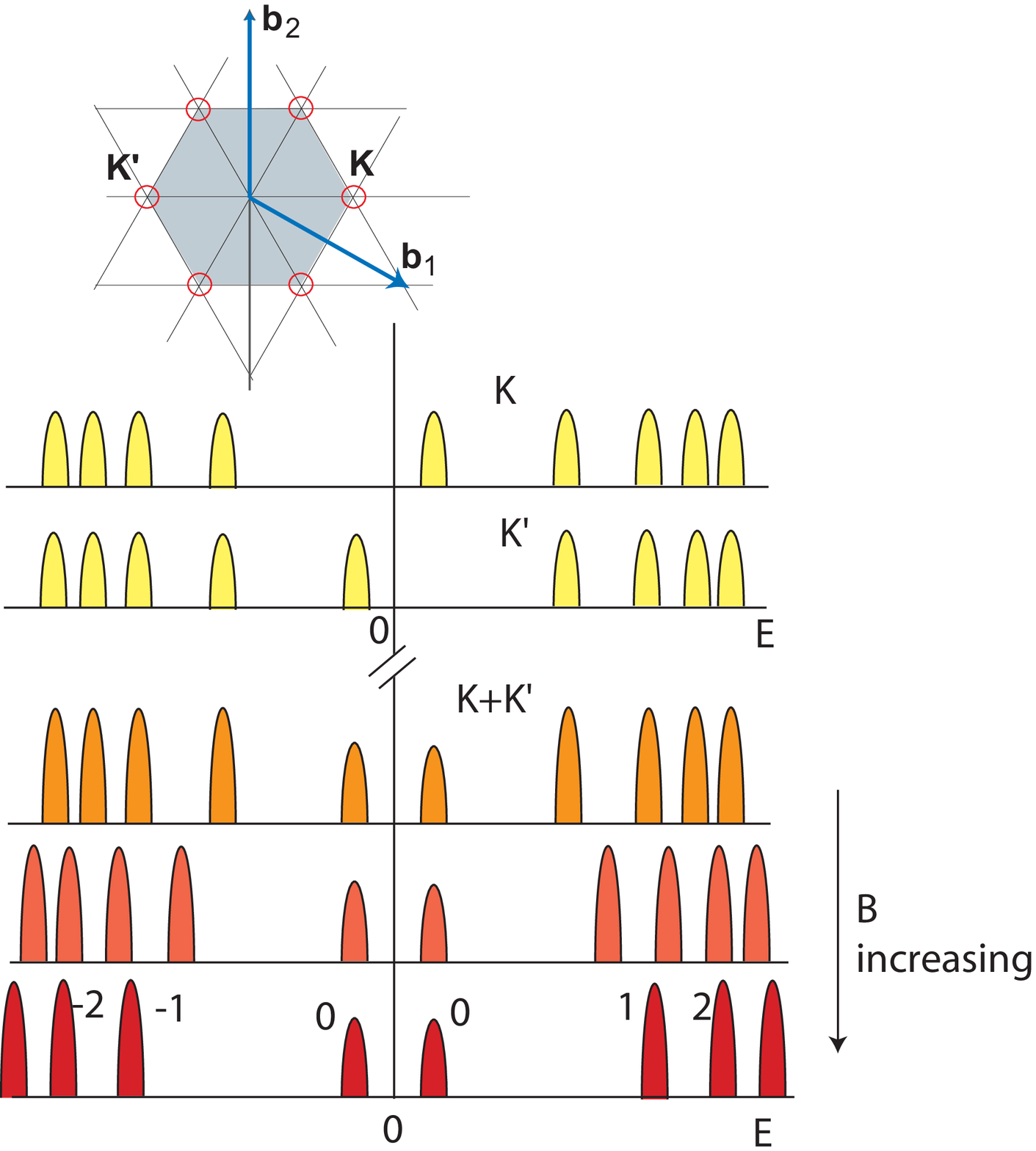}
\caption{\label{figspectrum}
The massive Dirac spectrum in the 2D hexagonal lattice. 
We assume distinct on-site energies on sublattices A and B (as in boron nitride). The Dirac cones sit at the high-symmetry points
{\bf K} and {\bf K'} on the edge of the Brillouin Zone (upper inset).
In a magnetic field $\bf B$, the $N=0$ LL at {\bf K} ({\bf K'}) is shifted up (down) from $E=0$ (upper curves).
The sum of the two spectra {\bf (K+K')} is symmetric about $E=0$. However, the $N=0$ LLs are non-degenerate whereas all
other LLs ($N\ne 0$) have a valley degeneracy of 2. As $B$ increases, all LLs fan out except for the $N=0$ levels.
}
\end{figure*}

\vspace{3mm}\noindent
{\bf  A simple example of massive Dirac spectrum}
An example illustrating the non-degeneracy of the $N=0$ LL is the spinless fermion on the 2D hexagonal lattice
(valley degeneracy replaces spin degeneracy in this example).
The sublattices A and B have distinct on-site energies $\epsilon_A$ and $\epsilon_B$ as in BN. The Dirac cones remain centered at the
inequivalent ``valleys'' {\bf K} and {\bf K'} in $\bf k$ space (inset, Fig. \ref{figspectrum}). Both valleys acquire a mass gap.

For states close to the valley at {\bf K}, the 2D massive Dirac Hamiltonian is 
\begin{eqnarray}
{\cal H}_{2D} = 
\left[  
\begin{array}{cc}
m 			&  k_x-ik_y\\
k_x+k_y  &  -m 
													\end{array} \right],
\label{H2D}
\end{eqnarray}
in the basis $(1,0)^T$ (pseudospin up) and $(0,1)^T$ (pseudospin down),
where {\bf k} is measured from $\bf K$ and $m>0$ represents the gap parameter proportional to $\epsilon_A-\epsilon_B$
(we set the velocity $v$ to 1). In a field $\bf B$, we replace $\bf k$ by $\boldsymbol{\pi} = {\bf k}-e{\bf A}$ with the vector gauge ${\bf A} = (0, Bx, 0)$. 
Introducing the operators
\be
a^\dagger = (\ell_B/\sqrt2) \pi_-, \quad a = (\ell_B/\sqrt2) \pi_+ ,
\label{aa}
\ee
with $\pi_{\pm} = \pi_x \pm i\pi_y$, and eigenstates $|N\rangle$ satisfying 
\be
a^\dagger |N\rangle = \sqrt{N+1}|N+1\rangle, \quad a |N\rangle = \sqrt{N}|N-1\rangle,
\label{an}
\ee
we diagonalize the Hamiltonian to get eigenenergies $E_N$ given by
\be
E_N^2 = {m^2 + (2N/{\ell_B^2})}
\label{EN}
\ee
(for brevity, we will write $E$ for $E_N$).

For positive $E$, the (unrenormalized) 2-spinor eigenstates are (for $N=0,1,\cdots$)
\begin{eqnarray}
|\Psi_{N,+}\rangle = \left(  
\begin{array}{c}
|N\rangle \\
\frac{1}{E+m}\frac{\sqrt{2N}}{\ell_B} |N-1\rangle
\end{array} \right), \quad (E>0).
\label{Psin}
\end{eqnarray}

For the negative energy states, the corresponding eigenvectors are ($N=1,2,\cdots$)
\begin{eqnarray}
|\Psi_{N,-}\rangle = \left(  
\begin{array}{c}
|N\rangle \\
-\frac{1}{|E|-m}\frac{\sqrt{2N}}{\ell_B} |N-1\rangle
\end{array} \right), \quad (E<0).
\label{Psin-}
\end{eqnarray}

Setting $N=0$ in Eq. \ref{Psin}, we find that the positive-energy state $|\Psi_{0,+}\rangle = (|0\rangle,0)^T$
(pseudospin up). For $E<0$, however, the lower entry in Eq. \ref{Psin-} is non-determinate (0/0).
This implies that the state $N=0$ does not exist for $E<0$. Thus, for the valley at {\bf K},
there is only one LL with $N=0$. It has positive energy $E_0 = |m|$; the corresponding LL at 
$-|m|$ is absent (we sketch the spectrum as the uppermost curve K in Fig. \ref{figspectrum}). 

Repeating the calculation for {\bf K'}, we find the opposite situation (the Hamiltonian is the conjugate of Eq. \ref{H2D}). 
Now the $N=0$ LL has energy $E_0 = -|m|$, but the $N=0$ LL is absent in the positive spectrum (curve K' in Fig. \ref{figspectrum}).

A transport experiment detects the sum of the two spectra ({\bf K+K'} in Fig. \ref{figspectrum}).
In the total spectrum, the two $N=0$ LLs are non-degenerate whereas all LLs with $N\ne 0$ have a valley degeneracy of 2.
The difference simply reflects the conservation of states.
In the limit $m\to 0$, we recover the spectrum of graphene. If, at finite $m$, each of the $N=0$ LLs had a valley degeneracy of 2, 
we would end up with an $N=0$ LL in graphene with 4-fold valley degeneracy.

The authors in Refs.~\cite{Wolff,Phillips,Bernevig} and others have shown that the non-degeneracy of the $N$=0 LL
also holds in massive Dirac systems even when a Rashba term and a Zeeman energy term are included.

\newpage

\newpage
\noindent
\vspace{5mm}\\
{\bf Acknowledgements}\\
We acknowledge helpful discussions with B. A. Bernevig, F. D. M. Haldane and M.Z. Hasan. 
The research is supported by the Army Research Office (ARO W911NF-11-
1-0379) and the US National Science Foundation (Grant No. DMR
0819860). T.L acknowledges scholarship support from the Japan Student Services Organization. 
High-field measurements were performed at the National High Magnetic Field Laboratory
which is supported by NSF (Award DMR-084173), by the State of Florida, and
by the Department of Energy. 
\vspace{5mm}\\
{\bf Author contributions}\\
T.L., Q.G., R.J.C. and N.P.O. planned and carried out the experiment. T.L. and N.P.O. analysed the data
and wrote the manuscript. J.X., M.H. and S.P.K. assisted with the measurements and analyses.  Q.G. and R.J.C. grew the crystals.
All authors contributed to editing the manuscript.
\vspace{5mm}\\
{\bf Additional information}\\
Supplementary information is available in accompanying article.
Correspondence and requests for materials should be addressed to T.L., Q.G., R.J.C. and N.P.O.
\vspace{5mm}\\
{\bf Competing financial interests}\\
The authors declare no competing interests.
\end{document}